# Positive time fractional derivative

W. Chen

Simula Research Laboratory, P. O. Box. 134, 1325 Lysaker, Norway

(30 September 2002)

## Abstract

In mathematical modeling of the non-squared frequency-dependent diffusions, also known as the anomalous diffusions, it is desirable to have a positive real Fourier transform for the time derivative of arbitrary fractional or odd integer order. The Fourier transform of the fractional time derivative in the Riemann-Liouville and Caputo senses, however, involves a complex power function of the fractional order. In this study, a positive time derivative of fractional or odd integer order is introduced to respect the positivity in modeling the anomalous diffusions.

## 1. Introduction

The time fractional derivative has long been found to be a very effective methodology to describe the anomalous attenuation (damping, diffusion) behaviors such as the arbitrary frequency-dependent damping (Makris and Constantinou, 1991; Enelund, 1996; Adhikari, 2000) and non-square frequency-dependent Burgers equation (Ochmann and Makarov, 1993). Caputo (1967) and Caputo and Mainardi (1971) developed the constant $Q$ seismic wave propagation model via the fractional derivative, where $Q$ represents an attenuation per unit wavelength. The Bagley-Torvik strain-stress model (Bagley and Torvik, 1983) has also been widely noticed for polymer solids without cross linking, where the fractional derivative is an essential building block. A fractional derivative time model without space variables can be typically expressed as (Seredynska and Hanyga, 2000)

$$D^2 u + \gamma D^{1+\eta} u + F(u) = 0, \tag{1}$$

where $D^{1+\eta}$ with $0<\eta<1$ represents the $(1+\eta)$-order fractional derivative with respect to time in the sense of Caputo (1967) and models the anomalous damping, and $\gamma$ is the thermoviscous coefficient. The detailed definition of the fractional derivative will be given in later section 2. Gaul (1999) points out that replacing the integer time derivative in the conventional PDE attenuation modeling by the fractional time derivative leads to better accuracy with fewer parameters and guaranteed causality.

The Fourier transform of the time fractional derivative in the sense of Riemann-Liouville or Caputo is

$$FT^+[D^\eta u] = (-i\omega)^\eta U, \tag{2}$$

where $FT^+$ denotes the forward Fourier transform, $\omega$ represents the angular frequency, and $U$ is the Fourier transform of the physical signal $u(t)$ such as the pressure or the velocity. The complex power function of fractional or odd order $\eta$ in (2) causes the difficulty in the analysis of the dispersion equation corresponding to the model equation (1) in terms of causality. It is also noted that if $\eta$ is an odd integer, we also face the same difficulty as a fraction. The purpose of this note is to introduce a positive time fractional derivative to overcome this troublesome issue.

## 2. Riemann-Liouville and Caputo fractional derivatives

The Riemann-Liouville fractional integral is an essential concept to understand the fractional derivative and is given by (Samko et al, 1987; Diethelm, 1997 and 2000)

$$J^q\{\psi(t)\} = \frac{1}{\Gamma(q)} \int_a^t \frac{\psi(\tau)}{(t-\tau)^{1-q}} d\tau, \qquad 0<q, \tag{3}$$

where $q$ and $a$ are real valued. The corresponding Riemann-Liouville fractional derivative is interpreted as

$$D_*^\lambda\{\psi(t)\} = \frac{d}{dt}\left[J^{1-\lambda}\{\psi(t)\}\right], \qquad 0<\lambda<1, \qquad (4)$$

which can be further elaborated into

$$\begin{aligned}D_*^\lambda\{\psi(t)\} &= \frac{1}{\Gamma(1-\lambda)}\frac{d}{dt}\int_a^t \frac{\psi(\tau)}{(t-\tau)^\lambda}d\tau \\ &= \frac{1}{\Gamma(-\lambda)}\int_a^t \frac{\psi(\tau)}{(t-\tau)^{1+\lambda}}d\tau \\ &= J^{-\lambda}\{\psi(t)\}.\end{aligned} \qquad (5)$$

The Riemann-Liouville fractional derivative has some notable disadvantages in engineering applications such as the hyper-singular improper integral, where the order of singularity is higher than the dimension, and nonzero of the fractional derivative of constants, e.g., $D^\lambda 1 \neq 0$, which would entail that dissipation does not vanish for a system in equilibrium (Samko et al, 1987; Seredynska and Hanyga, 2000) and invalidates the causality. The Caputo fractional derivative is instead developed to overcome these drawbacks (Caputo, 1967; Caputo and Mainardi, 1971) as defined below

$$D^\lambda\{\psi(x)\} = J^{1-\lambda}\left[\frac{d}{dx}\psi(x)\right]. \qquad (6)$$

Integration by part of (6) will lead to

$$\begin{aligned}D^\lambda\{\psi(t)\} &= \frac{1}{\Gamma(1-\lambda)}\int_a^t \frac{1}{(t-\tau)^\lambda}\frac{d\psi(\tau)}{d\tau}d\tau \\ &= D_*^\lambda\{\psi(t)\} - \frac{\psi(a)}{\Gamma(1-\lambda)}\frac{1}{(t-a)^\lambda}.\end{aligned} \qquad (7)$$

It is observed that the second term in (7) regularizes the Caputo fractional derivative to avoid the potentially divergence from singular integration at $t=a$. In addition, the Caputo fractional differentiation of a constant results in zero. The fractional

derivatives in the Riemann-Liouville and Caputo senses can respectively be in general expressed as

$$D_*^\mu \{\psi(x)\} = D^m [J^{m-\mu} \psi(x)], \qquad (8)$$

$$D^\mu \{\psi(x)\} = J^{m-\mu}[D^m \psi(x)]. \quad m-1 \prec \mu \leq m, \qquad (9)$$

where $m$ is an integer, and $D^m \psi = d^m \psi / dt^m$.

### 3. Positive time fractional derivative

The positivity of the attenuation (damping) operation required by the decay of the global energy (Matignon et al, 1998) can well be preserved via a positive fractional calculus operation. The space fractional Laplacian (Hanyga, 2001; Chen and Holm, 2002) is the positive definition operator and thus very suitable to describe the anomalous attenuation behaviors, while the traditional time fractional derivative lacks this crucial positive property as shown in the Fourier transform (2). Lighthill (1962) gave the following positive real power frequency functions of non-integer order $\eta$

$$FT^-\left(|\omega|^\eta\right) = \Gamma(\eta+1)\cos[(\eta+1)\pi/2]/\left(\pi|t|^{\eta+1}\right) \qquad (10a)$$
$$= s(t)$$

and of non-integer order $\eta$

$$FT^-\left(\omega^\eta \operatorname{sgn}(\omega)\right) = \Gamma(\eta+1)(-1)^{(\eta+1)/2}/\left(\pi|t|^{\eta+1}\right) \qquad (10b)$$
$$= s(t),$$

where $FT^-$ is the inverse Fourier transform operation. (10) is crucial for the new lossy acoustic molds developed by Szabo (1994). It is noted that (10b) can be covered by (10a). In terms of (10), we can introduce a positive time fractional derivative, whose Fourier transform characterizes the positive operation as follows

$$FT^+\left(D^{|\eta|}u\right)=|\omega|^\eta U, \tag{11}$$

where $FT^+$ is the Fourier transform operation, and

$$D_*^{|\eta|}u = s(t)*u(t) = \frac{1}{q(\eta)}\int_0^t \frac{u(\tau)}{(t-\tau)^{\eta+1}}d\tau, \tag{12}$$

$$D^{|\eta|}u = \begin{cases} \dfrac{1}{\eta q(\eta)}\int_0^t \dfrac{D^1 u(\tau)}{(t-\tau)^\eta}d\tau, & 0 \prec \eta \le 1, \\ \dfrac{1}{\eta(\eta-1)q(\eta)}\int_0^t \dfrac{D^2 u(\tau)}{(t-\tau)^{\eta-1}}d\tau, & 1 \prec \eta \prec 2, \end{cases} \tag{13}$$

where * denotes the convolution product operation, and

$$q = \frac{\pi}{\Gamma(\eta+1)\cos[(\eta+1)\pi/2]}. \tag{14}$$

The new definitions (12) and (13) of the positive fractional derivative are to combine the Fourier transform relationship (10) respectively with the Riemann-Liouville fractional derivative (8) and with the Caputo fractional derivative (9) to hold the real positive Fourier transform relationship (11).

The positive fractional derivative (13) can further be generalized by

$$D^{|\sigma|}u = \begin{cases} \dfrac{1}{(\sigma-2k)q(\sigma)}\int_0^t \dfrac{D^{2k+1}u(\tau)}{(t-\tau)^{\sigma-2k}}d\tau, & 2k \prec \sigma \le 2k+1, \\ \dfrac{1}{(\sigma-2k)[\sigma-(2k+1)]q(\sigma)}\int_0^t \dfrac{D^{2k+2}u(\tau)}{(t-\tau)^{\sigma-(2k+1)}}d\tau, & 2k+1 \prec \sigma \prec 2k+2, \end{cases} \tag{15}$$

where $k$ is a non-negative integer. We also have

$$D^{|\eta|+l}u = D^{|\eta|}D^l u \ne D^l D^{|\eta|}u, \tag{16}$$

$$D_*^{|\eta|+l}u = D^l D_*^{|\eta|}u \neq D_*^{|\eta|}D^l u, \qquad (17)$$

where $l$ is a positive integer number. Note that $D^{|2k+1|}u \neq D_*^{|2k+1|}u \neq D^{2k+1}u$. It is straightforward to have

$$FT^+\left(D^{|\eta|+l}u\right) = (-i\omega)^l |\omega|^\eta U. \qquad (18)$$

**References**

Adhikari, S. (2000), *Damping Models for Structural Vibration*, Ph.D. thesis, Cambridge University.

Baglegy, R. L. and Torvik, P. J. (1983), "A theoretical basis for the application of fractional calculus to viscoelasticity," *J. Rheol*. **27**, 201-210.

Caputo, M. (1967), "Linear models of dissipation whose $Q$ is almost frequency independent-II," *Goephys. J. R. atr. Soc.*, **13**, 529-539.

Caputo, M. and Mainardi, F. (1971), "A new dissipation model based on memory mechanism," *Pure and Appl. Geophys*. **91**, 134-147.

Chen, W. and Holm, S. (2002), "Fractional Laplacian, Lévy stable distribution, and time-domain models for linear and nonlinear frequency-dependent lossy media," *Ultrasound project report*, Simula Research Lab.

Diethelm, K. (1997), "An algorithm for the numerical solution of differential equations of fractional order," *Electronic Trans. Numer. Anal*. **5**, 1-6.

Diethelm, K. (2000), *Fractional Differential Equations, Theory and Numerical Treatment*, preprint.

Enelund, M. (1996), *Fractional Calculus and Linear Viscoelasticity in Structural Dynamics*, Ph.D thesis, Chalmers University of Technology, Sweden.

Gaul, L. (1999), "The influence of damping on waves and vibrations," *Mechanical Systems and Signal Processing*, **13**(1), 1-30.

Hanyga, A. (2001), "Multi-dimensional solutions of space-fractional diffusion equations," *Proc. R. Soc. London A,* **457**, 2993-3005.

Lighthill, M. J. (1962), *Introduction to Fourier Analysis and Generalized Functions* (Cambridge UP, Cambridge).

Makris, N and Constantinou, M. C. (1991), Fractional-derivative Maxwell model for viscous dampers, *J. Struct. Engng.*, **117**(9), 2708-2724.

Matignon, D., Audounet, J. and Montseny, G. (1998), "Energy decay for wave equations with damping of fractional order," *LAAS Report 98031*.

Ochmann, M. and Makarov, S., (1993), "Representation of the absorption of nonlinear waves by fractional derivative," J. Acoust. Soc. Am. **94**(6), 3392-3399.

Samko, S. G., Kilbas, A. A., Marichev, O. I. (1987), *Fractional Integrals and Derivatives: Theory and Applications* (Gordon and Breach Science Publishers).

Seredynska, M. and Hanyga, A. (2000), "Nonlinear Hamiltonian equations with fractional damping," *J. Math. Phys.*, **41**, 2135-2156.

Szabo, T. L. (1994), "Time domain wave equations for lossy media obeying a frequency power law," *J. Acoust. Soc. Amer*. **96**(1), 491-500.